# "Medium-tech" industries may be of greater importance to a local economy than "High-tech" firms:

**New methods for measuring the knowledge base of an economic system**

Wilfred Dolfsma[1] & Loet Leydesdorff[2]

**Abstract.** In this paper we offer a way to measure the knowledge base of an economy in terms of probabilistic entropy. This measure, we hypothesize, is an indication of the extent to which a system, including the economic system, self-organizes. In a self-organizing system, interactions between dimensions or subsystems will unintentionally give rise to anticipations that are properly aligned. The potential reduction of uncertainty can be measured as negative entropy in the mutual information among three (or more) dimensions. For a knowledge-based economy, three dimensions can be considered as key: the distribution of firm sizes, the geographical locations, and the technological classifications of firms. Based on statistics of these three dimensions and drawing on a unique dataset of all Dutch firms registered with the Chambers of Commerce, we are able to refine well-known empirical findings for the geographical dimension. Counter-intuitive, however, are our empirical findings for the dimension of technology. Knowledge diffusion through medium-tech industry is much more important for a localized economy than knowledge creation in high-tech industry. Knowledge-intensive services tend to uncouple economic activities from the regional dimension.

Scholars and policy makers have increasingly recognized the contribution of knowledge to the economy and to economic growth. However, how exactly intangibles like knowledge can make these contributions to the economy is not clear. This lack of clarity is partly due to the problem of *measuring* the contribution which knowledge makes to the dynamics of an economy [1], and has led some scholars to make unsubstantiated claims about how certain sectors and groups of firms contribute to the knowledge economy. A lack of clarity may lead policy makers to unfounded and possibly misguided decisions [2].

---

[1] University of Groningen School of Economics and Business, PO Box 800, 9700 AV Groningen, The Netherlands; w.a.dolfsma@rug.nl.
[2] Amsterdam School of Communications Research (ASCoR), University of Amsterdam, Kloveniersburgwal 48, 1012 CX Amsterdam, The Netherlands; loet@leydesdorff.net; http://www.leydesdorff.net.

Using information theory [3, 4], we submit that the knowledge base of an economy can be measured. The measure suggested can be disaggregated to indicate the most significant contributions to the knowledge-based economy. Thus, we offer both a significant methodological contribution as well as an empirical contribution that is relevant for both scholars and policy makers. As the method used in this paper can be and has been used in other empirical contexts as well [3], perhaps the most significant contribution of this line of work is methodological.

On the basis of a *complete* set of Dutch firms, registered at the Chambers of Commerce, we calculate which region or type of industry contributes to the knowledge economy in terms of adding negative entropy to the Netherlands as a system of innovations. Negative entropy is a measure suggested in information science for the extent to which a system is self-organized. In a self-organized system, uncertainty is reduced autonomously as anticipations are harmonized between subsystems. If a system shows no negative entropy, interactions among its main dimensions or subsystems are disconnected and do not inform each other. The system thus needs to deal with the resulting uncertainty and will tend to disintegrate.

Our most striking empirical findings pertain to one of the three dimensions that are relevant for a knowledge economy: the technological dimension. Contrary to widely held expectations, medium-tech industries contribute far more to the knowledge base of an economy than high-tech manufacturing sectors. Furthermore, the much-heralded knowledge-intensive services do not add to the knowledge base of an economy. Only the high-tech end of these services contributes as much as but not more than the average.

**Information Theory: Methods and Data**

Insights from particularly thermodynamics, nonlinear dynamics, scientometrics, and information theory, allow us to measure this 'intangible' entity as a latent structure that can reduce uncertainty in a system in some configurations more than others [5]. Using probabilistic entropy statistics [6], one can measure the mutual information in three dimensions of a system [4]. Information flows among three dimensions can result in negative Shannon-type entropy or reduction of uncertainty at the systems level. This local reduction of uncertainty can be considered as an indicator of self-organization in a system [7, 8]. The emerging complex patterns are not the result of any of the specific subsystems or their relations, but 'configurational' and can thus be measured empirically [9]. In a self-organizing system, purposeful behaviour by actors can be expected to lead to unintended consequences insofar as the system's dynamics is determined more by internal dynamics than by disturbances. The capacity to generate negative entropy implies that disturbances can to some extent be filtered out by the system as noise [10], allowing a system without outside interventions to achieve orderly development along a stable path. Alignment of expectations is endogenous to this self-organization of the knowledge-based system.

| High tech Manufacturing | Knowledge-intensive sectors (KIS) |
|---|---|
| **30** Manufacturing of office machinery and computers<br>**32** Manufacturing of radio, television and communication equipment and apparatus<br>**33** Manufacturing of medical precision and optical instruments, watches and clocks<br><br>*Medium-high tech manufacturing*<br><br>**24** Manufacture of chemicals and chemical products<br>**29** Manufacture of machinery and equipment n.e.c.<br>**31** Manufacture of electrical machinery and apparatus n.e.c.<br>**34** Manufacture of motor vehicles, trailers and semi-trailers<br>**35** Manufacturing of other transport equipment | **61** Water transport<br>**62** Air transport<br>**64** Post and telecommunications<br>**65** Financial intermediation, except insurance and pension funding<br>**66** Insurance and pension funding, except compulsory social security<br>**67** Activities auxiliary to financial intermediation<br>**70** Real estate activities<br>**71** Renting of machinery and equipment without operator and of personal and household goods<br>**72** Computer and related activities<br>**73** Research and development<br>**74** Other business activities<br>**80** Education<br>**85** Health and social work<br>**92** Recreational, cultural and sporting activities<br><br>Of these **64, 72** and **73** are *high tech services*. |

**Table 1**: Classification of sectors according to [11].

We obtained data for the complete set of 1,131,668 firms registered with the Netherlands Chambers of Commerce for the Spring of 2001. This data is unique both for its completeness and for its level of detail. For each organization (firm) in the database, information on three dimensions was gathered: geography (*g*), technology (*t*), and organization (*o*). These are dimensions that are of fundamental importance for the knowledge-based economy [12]. We use the following indicators for these three dimensions: postal code for geographical location, firm size for organization, and finally sector code for the technology [13, 14; see Table 1]. Relevant sector codes for the medical sector are: **33** "Manufacturing of medical precision and optical instruments, watches and clocks," and **85** "Health and social work." Other relevant sectors would appear at a more detailed level of industry classification. Larger firms are structured differently from smaller ones, which has particular consequences for how information and knowledge are exchanged. In a related study [15] we discussed in more detail why these are appropriate indicators.

Depending on the empirical configuration, dimensions form a structure that allows communication and exchange of information to varying degrees. Probabilistic entropy as developed in information theory offers the possibility to analyze this configurational information or transmission between dimensions of a system that is mathematically manageable [7, 8]. Complex patterns can emerge when three dimensions interact, and uncertainty in a system consequently can either increase or decrease [16]. The negative entropy can also be considered as synergy which results from the fine-tuning of patterns of relations among different dimensions.

According to Shannon [17] the uncertainty in a probability distribution in a dimension (for example, in *t* for the distribution of technological classifications) can be formulated as follows:

$$H_t = - \sum_t p_t \log_2 p_t \qquad (1)$$

The mutual information between two variables *t* and *o* is given by:

$$T_{to} = H_t + H_o - H_{to} \qquad (2)$$

where $H_{to} = -\sum_t \sum_o p_{to} \log_2 p_{to}$. $T_{to}$ can be also be considered a measure of co-variance [18]. By taking the $\log_2$, values are expressed in bits of information. Mutual information among three dimensions provides the entropy score for a system as a whole using the following formula [8]:

$$T_{gto} = H_t + H_o + H_g - H_{to} - H_{gt} - H_{go} + H_{gto} \qquad (3)$$

Note that bilateral relations between dimensions reduce uncertainty, while interaction among three dimensions ($H_{gto}$) increases uncertainty and unfavourably affect entropy scores. Negative entropy scores indicate synergy within the system, a sign for the likelihood of information being exchanged—a favourable sign indicating self-organization in such a system.

**Findings**

The entropy value $T$ for the degree of self-organization within a system as a whole—and within each of its subsystems—can thus be determined empirically. Decomposition of the uncertainty into the various dimensions is possible because all Shannon formulas are based on aggregations (Σs) [3, 6, 19].

The Netherlands as a whole has a negative entropy score (-34 mbits), a favourable sign. Within its knowledge base uncertainty is reduced as expectations are mutually aligned. Decomposing[3] for the geographical dimension—calculations not presented here for lack of

---
[3] Decomposition can alter the scale of findings so that, for instance, scores in Table 2 and in Figure 1 are not immediately comparable.

space—we found results similar to those in other studies using methodologies more common in economics, with some notable exceptions, which we interpret as a corroboration of both the method and findings in this paper. What is striking, for example, is that the capital, Amsterdam, stands out even more when compared to its surrounding areas according to our method of analysis than it does in more standard location factor analyses that economic geographers tend to do [20]. Despite a dearth of high-tech firms in the larger area surrounding Rotterdam and its large port, in our analysis it comes out stronger than in conventional analyses. The south-east Limburg province as well as the north-east Groningen province come out less strongly.

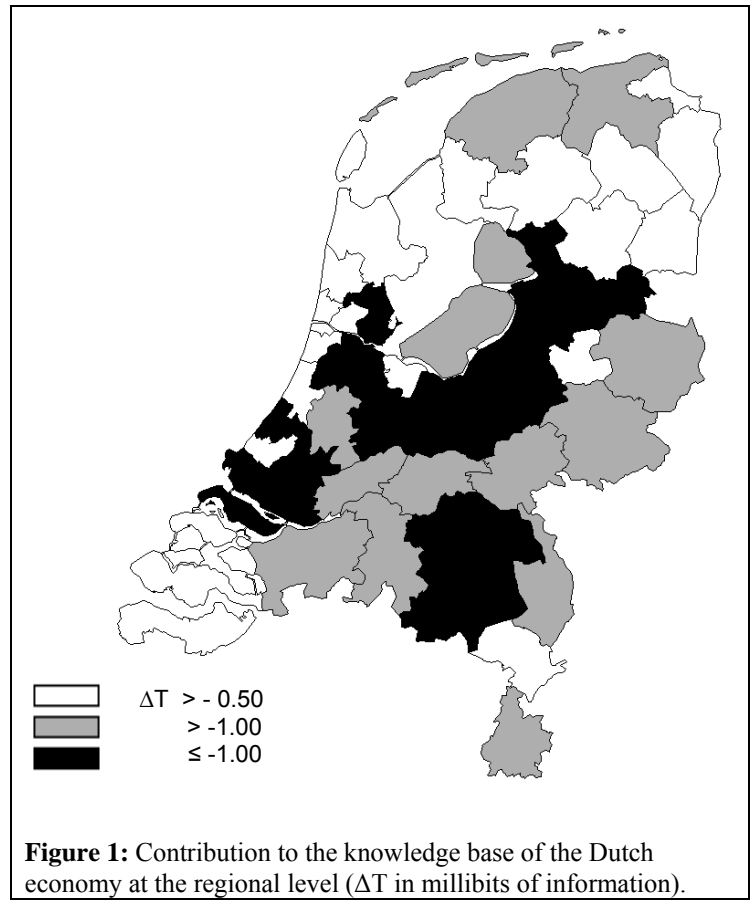

**Figure 1:** Contribution to the knowledge base of the Dutch economy at the regional level (ΔT in millibits of information).

Some decompositions in the dimension of sector / technology are provided in Table 2. High-tech manufacturing contributes favourably to the negative entropy of the system ($T = -60$ mbits; not shown in Table 2). Contrary to expectations however, the contribution of medium-tech industries is more important by an order of magnitude (after normalization): the combined score for medium & high-tech manufacturing is -219 mbits in comparison to the complete set. This finding is confirmed in other studies using this approach [21, 22].

As one expects high-tech sectors to be involved in knowledge creation and medium-tech sectors in knowledge diffusion, our results indicate the importance specifically of knowledge diffusion within a national or regional economy. The relatively much larger contribution to the knowledge base of an economy made by medium-tech firms also indicates the extent to which these firms are related to their immediate surroundings in terms of other firms, customers, and employees, while high-tech firms tend to be more linked globally and thus less embedded locally.

| All Sectors | | High- & Medium-tech Manufacturing | | | Knowledge-intensive Services | | | High-tech Services | | |
| --- | --- | --- | --- | --- | --- | --- | --- | --- | --- | --- |
| $T_{gto}$ | N | $T_{gto}$ | %‡ | N | $T_{gto}$ | %‡ | N | $T_{gto}$ | %‡ | N |
| -34 | 1,131,668 | -219 | -544 | 13,422 | -24 | +27.3 | 581,196 | -34 | 0 | 41,002 |

**Table 2**: Probabilistic entropy scores, by type of sector (in mbits of information).
‡ Percentage change compared to the $T_{gto}$ value for 'All Sectors'.

Knowledge-intensive services, including for instance technical consultancies as well as "health and social work," are often believed to diffuse knowledge among firms in an economy. However, these services are also relatively independent of their specific locations. The vicinity of an airport or railway station, that is, mobility, can be more important than the local conditions. In terms of the regional indicators and at the national level of the Netherlands, knowledge-intensive services do not reduce uncertainty and thus make no apparent contribution to the knowledge base of the economy. Their entropy score (-24 mbits) is not as favourable as the overall one: knowledge-intensive services increase uncertainty and disrupt the process of locally aligning expectations when compared to the average. The high-tech end of the knowledge-intensive services (e.g., R&D facilities) decreases uncertainty, but not more than the average for the national economy (-34 mbits). It is noteworthy that similar results were found for Germany [21].[4]

While sectors related to the medical subsystem of an economy (numbers 33 and 85) may be strongly involved in generating new knowledge, as indicated by the number of patents in these sectors [23], these high-tech contributions may matter more for a global economy than for a local (e.g., Dutch) one.

**Policy implications**

Elaborating upon information theory [4] we show that the knowledge base of an economy can be measured. Our contribution will allow the academic and the policy discussion to attain a degree of focus and be based on firmer ground. In addition, we actually measured the contributions of different types of sectors as well as regions to the knowledge base of an economy as a complex system. The analysis along the geographical dimension does not provide surprising insights. However, the analysis of the technological or sectoral dimension does.

---

[4] In Germany, the effect of high-tech services counteracting on the uncoupling effect of knowledge-intensive services on the regional economy was larger, notably in East-Germany (the former GDR).

Medium-tech industries provide by far the most important contribution to the knowledge-based economy. One should not discourage high-tech manufacturing sectors, including manufacturing of medical precision and optical instruments (sector 33), based on these findings, but one should nevertheless realize that firms in these sectors both draw less extensively on and contribute less to regional knowledge bases. Knowledge-intensive services, though influencing the knowledge economy unfavourably at the regional level, will nevertheless be indispensable for an economy like the Dutch one if only because these firms make up half the total number of firms. Based on this analysis of what contributes to the knowledge base of the economy, using insights from information theory, a shift in the emphasis of government technology or innovation policy is thus suggested from a focus on high-tech to more focus on the large group of medium-tech firms, most of which are SMEs. As for regional development, one should be careful with the stimulation of knowledge-intensive services because economic diffusion of these services does not necessarily sediment in the region itself.